# Spin-squeezed vector atomic magnetometry

Jinyang Li[1], Gour Pati[2], Renu Tripathi[2], and Selim M Shahriar[1,3]

[1]*Department of Physics and Astronomy, Northwestern University, Evanston, IL 60208, USA*

[2] *Division of Physics, Engineering, Mathematics and Computer Science, Delaware State University, Dover, DE 19901, USA*

[3]*Department of Electrical and Computer Engineering, Northwestern University, Evanston, IL 60208, USA*

**Abstract**

Atomic magnetometers based on Zeeman shift measurement have the potential for high sensitivity and long-term stability. Like other atomic sensors including atomic clocks and atom interferometers, the atomic magnetometer could in principle be augmented with spin squeezing for further sensitivity enhancement. However, existing atomic magnetometers are not compatible with spin squeezing because the atoms can hardly be in a pure quantum state during operation. A natural challenge is the arbitrary direction of the magnetic field. In this paper, we propose a cold-atom-based magnetometer with spin squeezing that can measure both the magnitude and the direction of an arbitrary magnetic field. For experimentally accessible parameters, we show that the technique described above could achieve a sensitivity nearly three orders of magnitude higher than that of the best existing magnetometers.

## 1. Introduction

Precise magnetic field sensing is of great significance in fundamental physics, geology, navigation, biology, and medical science [1,2,3,4,5,6,7,8,9,10,11]. Atomic magnetometers sense a magnetic field by measuring Larmor precession or Zeeman shifts. For atomic magnetometers based on Larmor precession measurement [3,4,5,6,7,8,9,12], a pumping beam together with a probe beam is applied to the atoms when they are exposed to a radio-frequency field. If the radio-frequency field is resonant with the Larmor precession frequency, the absorptance, the phase, or the polarization of the probe beam will change significantly. In this way, the Lamor precession frequency is measured. It should be noted that magnetometers based on Lamor precession measurement typically cannot measure the direction of the magnetic field, and thus are not vector magnetometers. Ref. [13] proposes a way to apply measurement-induced spin squeezing to such an magnetometer. However, this proposed magnetometer even requires the magnetic field to align in a particular direction. Accordingly, it is not obvious how to make magnetometers based on Lamor frequency measurement spin-squeezed vector magnetometers. A prototypical atomic magnetometer employing the measurement of Zeeman shifts is based on coherent population trapping (CPT) [10,11,14,15,16]. During operation, the relative frequency between the two CPT beams is scanned. If the relative frequency matches a transition frequency between two Zeeman substates, with the quantization axis defined to be along the direction of the magnetic field, a CPT peak will occur. The distance between two adjacent CPT peaks indicates the magnitude of the magnetic field. By analyzing the relative amplitudes of the CPT peaks, it is also possible to determine the direction of the magnetic field. To illustrate the underlying principle, we note first that the CPT process occurs via optical pumping from the bright state to the dark state (assuming, for simplicity and without loss of generality, that the system consists of only one bright state and one dark state). The pumping rate is proportional to the square of the Rabi frequency of each leg of the relevant $\Lambda$ system. For a given choice of polarizations of the two laser fields, these Rabi frequencies would depend on the choice of the quantization axis (the direction of the magnetic field). In a system employing warm atomic vapor, the process of optical pumping into the dark state is countered by dephasing mechanism due to, for example, atomic collision. Thus, even in the steady state, the amplitudes of different CPT peaks vary due to unequal optical pumping rate for different CPT peaks. By deciphering the pumping rates via analyzing the CPT peaks, we can then determine the direction of the magnetic field (i.e., the quantization axis). It can be noticed that decoherence effects including collision and dephasing play an important role in the working principle. The presence of such decoherence effects makes it difficult to implement high-fidelity spin squeezing.

Consider next the prospect of realizing a CPT magnetometer employing cold atoms. In this case, the measurement of the magnitude of the magnetic field can be achieved in the same way as the one mentioned above for the case of atomic vapor. However, care must be taken to determine the direction of the magnetic field. Since in the ideal limit there is no collisional dephasing to counter the optical pumping process from the bright states to the dark states, the CPT peaks would all be equally high in the steady state. To avoid this problem, one can carry out measurements on a time scale shorter than the reciprocal of the optical pumping rate, in order to determine the optical pumping rate for each CPT transition. This information in turn can be used to determine the direction of the magnetic field. However, a constraint imposed by such a technique is that for each CPT transition, a fraction of the atoms will remain in the bright state when the measurement is performed. Because the existence of atoms in the bright state would correspond to mixed quantum states of the atoms, it would again not be possible to implement high-fidelity spin squeezing using such a scheme. In this paper, we propose an atomic magnetometer employing cold atoms that not only can measure both the magnitude and the direction of a magnetic field, but also is compatible with spin squeezing.

The rest of the paper is organized as follows. In Sec. 2, we describe a cold-atom-based vector atomic magnetometer employing atoms in a pure quantum state. In Sec. 3, we show how to augment the sensitivity of this magnetometer by employing spin-squeezing. Concluding remarks are presented in Sec. 4. The Appendix shows a simplified derivation of the squeezing parameter for the case of one-axis-twist squeezing.

## 2. Vector atomic magnetometer with atoms in a pure state

We first describe a vector atomic magnetometer employing cold atoms that stay in a pure state until detection. In what follows, we assume that the magnetic field to be measured is in the weak-field regime where the Zeeman shifts are much smaller than the hyperfine splitting. Of course, this assumption limits the dynamic range of the proposed magnetometer. However, for most applications of interest, the strength of the magnetic field to be measured is expected to fall well within this limit.

In what follows, the quantization axis, denoted as the $z$ axis, is defined as the direction of the magnetic field to be measured, so that $m_F$ is a good quantum number for the Hamiltonian in the absence of microwave or optical fields. For concreteness, $^{87}$Rb is used as an example for the illustration of such a magnetometer. The process would start with the atoms distributed among all the Zeeman substates in the $5S_{1/2}, F=2$ hyperfine level. At this point, the magnetic field and optical fields used for the magneto-optic trap (MOT) have been turned off. The next step is to concentrate all the atoms into the $5S_{1/2}, F=1, m_F=0$ Zeeman substate. The process is described below. A linearly polarized microwave pulse, tuned to the frequency resonant with the transition from $5S_{1/2}, F=2, m_F=0$ to $5S_{1/2}, F=1, m_F=0$, is first applied. The polarization direction [17] of the microwave field is not particularly chosen due to the ignorance of the orientation of the $z$ axis. For convenience in description, the polarization direction is denoted as the $z'$ axis.

Generally, the $z'$ axis does not coincide with the $z$ axis. As such, the microwave pulse comprises all three polarizations, namely $\pi$, $\sigma^+$, and $\sigma^-$, with the two circular polarizations ($\sigma^+$ and $\sigma^-$) equally weighted. Therefore, the atoms can be transferred to each of the following three states: $S_{1/2}, F=1, m_F \in \{-1,0,1\}$. These three transitions will in general have different resonant frequencies, with two adjacent resonant angular frequencies differing by $g_F \mu_B B/\hbar$, where $g_F$ is the gyromagnetic factor for this hyperfine state, $\mu_B$ is the Bohr magneton, and $B$ is the magnitude of the magnetic field. If this frequency gap is much larger than the transition linewidth [18], only one transition can occur efficiently for a certain frequency of the microwave field. To suppress off-resonant transitions, the microwave field should be weak enough to limit the power broadening, and Blackman pulses [19, 20, 21] instead of square pulses should be used. After this step, we apply an optical field to blow away the remaining atoms in the $5S_{1/2}, F=2$ state. As a result, all the atoms will be prepared in the $5S_{1/2}, F=1, m_F=0$ state.

Following the state preparation process is the actual measurement operation, which starts with the application of another microwave pulse linearly polarized along the $z'$ axis. The effective Rabi frequency (it is an effective Rabi frequency because for a Blackman pulse, the Rabi frequency is a function of time [19]) of this microwave pulse is denoted as $\Omega$ if $\theta = 0$, where $\theta$ is the angle between the $z$ and $z'$ axis. The effective duration of the pulse is set at $\tau = \pi/\Omega$. Since $\theta$ does not necessarily equal zero, the microwave pulse in this case would also comprises all three polarizations, namely $\pi$, $\sigma^+$, and $\sigma^-$, with the two circular polarizations ($\sigma^+$ and $\sigma^-$) equally weighted. Therefore, atoms can be excited to each of the following three states: $S_{1/2}, F=2, m_F \in \{-1,0,1\}$, as shown in Figure 1. Again, to suppress off-resonant transitions, the microwave field should be weak enough to limit the power broadening, and Blackman pulses instead of square pulses should be used.

Simulation results of the spectrum resulting from the three transitions with the application of Blackman pulses are shown in Figure 2. An analysis of the heights of the resonant peaks can be used to determine the value of $\theta$. As shown in Figure 2(a) and Figure 2(c), a single peak at the center would indicate $\theta = 0$ because there is only the $\pi$-polarized component, and no peak at the center would indicate $\theta = \pi/2$ because the $\pi$-polarized component is absent. For other values of $\theta$, all three peaks will be present. Noting that the Rabi frequency is $\Omega\cos\theta$ for the $\pi$- polarized component, and $(\Omega \sin\theta)/2$ for the $\sigma^{\pm}$-polarized component, the height of the central peak is given by $[1-\cos(\Omega\tau\cos\theta)]/2 = [1-\cos(\pi\cos\theta)]/2$, and the heigh of the side peaks is given by $\{1-\cos[(\Omega\tau\sin\theta)/2]\}/2 = \{1-\cos[(\pi\sin\theta)/2]\}/2$, recalling that $\Omega\tau = \pi$, as discussed above. Here, the unity height corresponds to the case where all atoms are in the detected state ( $F=2$ ). The case of $\theta = \pi/4$ is shown in Figure 2(b).

It can be shown that in this approach, microwave pulses with linear polarizations in three orthogonal directions are needed to uniquely determine the direction of the magnetic field. It should be noted that, in practice, horn antennas are normally used to radiate such a microwave field. These horn antennas can only radiate a microwave field with a linear polarization in a certain direction. Therefore, it would be necessary either to use three horn antennas, or rotate a single antenna to three different orientations for these measurements.

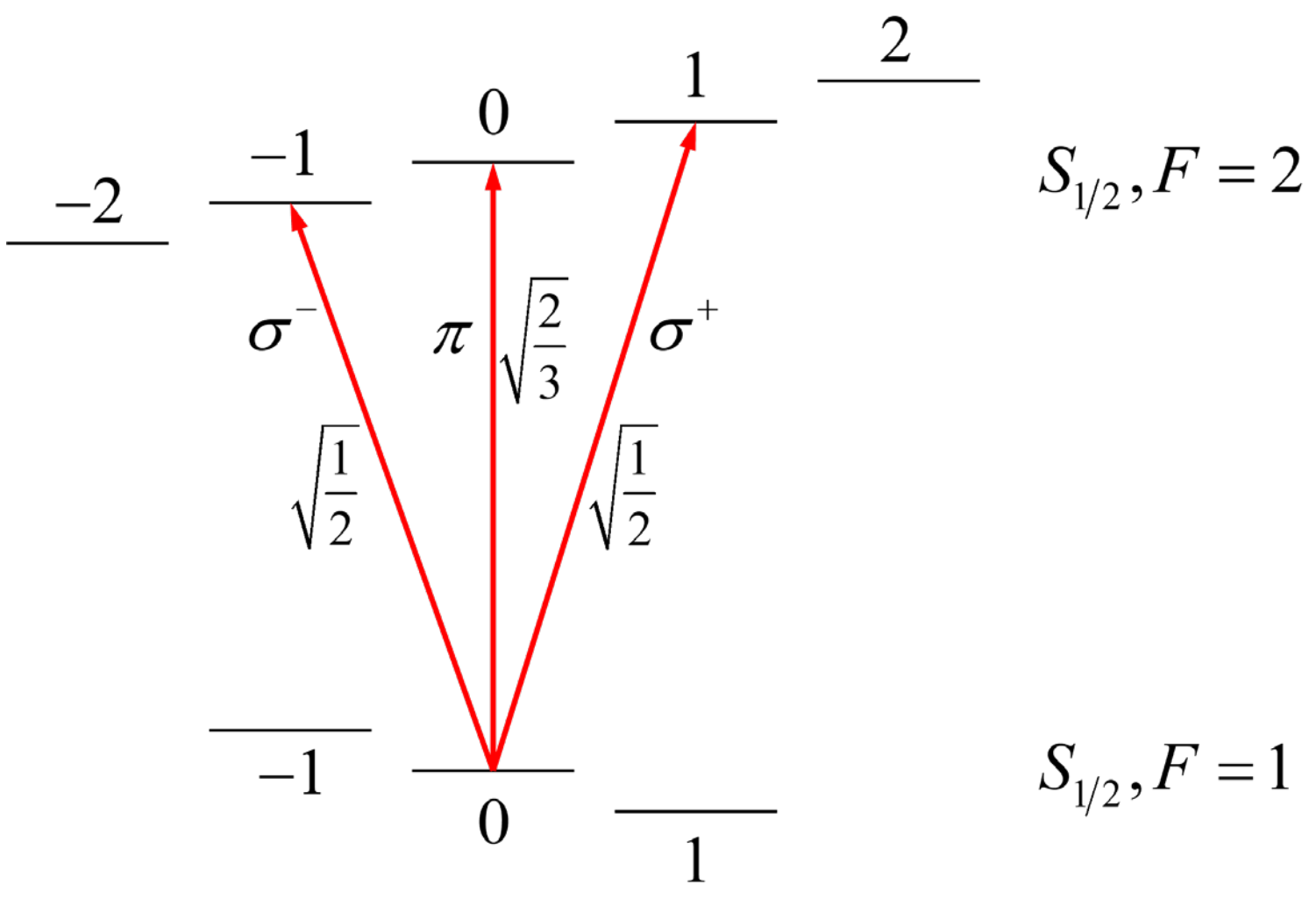

Figure 1. Transitions relevant for the vector atomic magnetometer using $^{87}$Rb. The matrix elements of the transitions are also shown. Atoms are initially prepared in the state $F=1, m_F=0$. A linearly polarized microwave pulse is applied to drive the transitions from $F=1$ to $F=2$. Because the direction of the magnetic field is arbitrary, the polarization of the microwave pulse could be a superposition of $\pi$, $\sigma^+$, and $\sigma^-$.

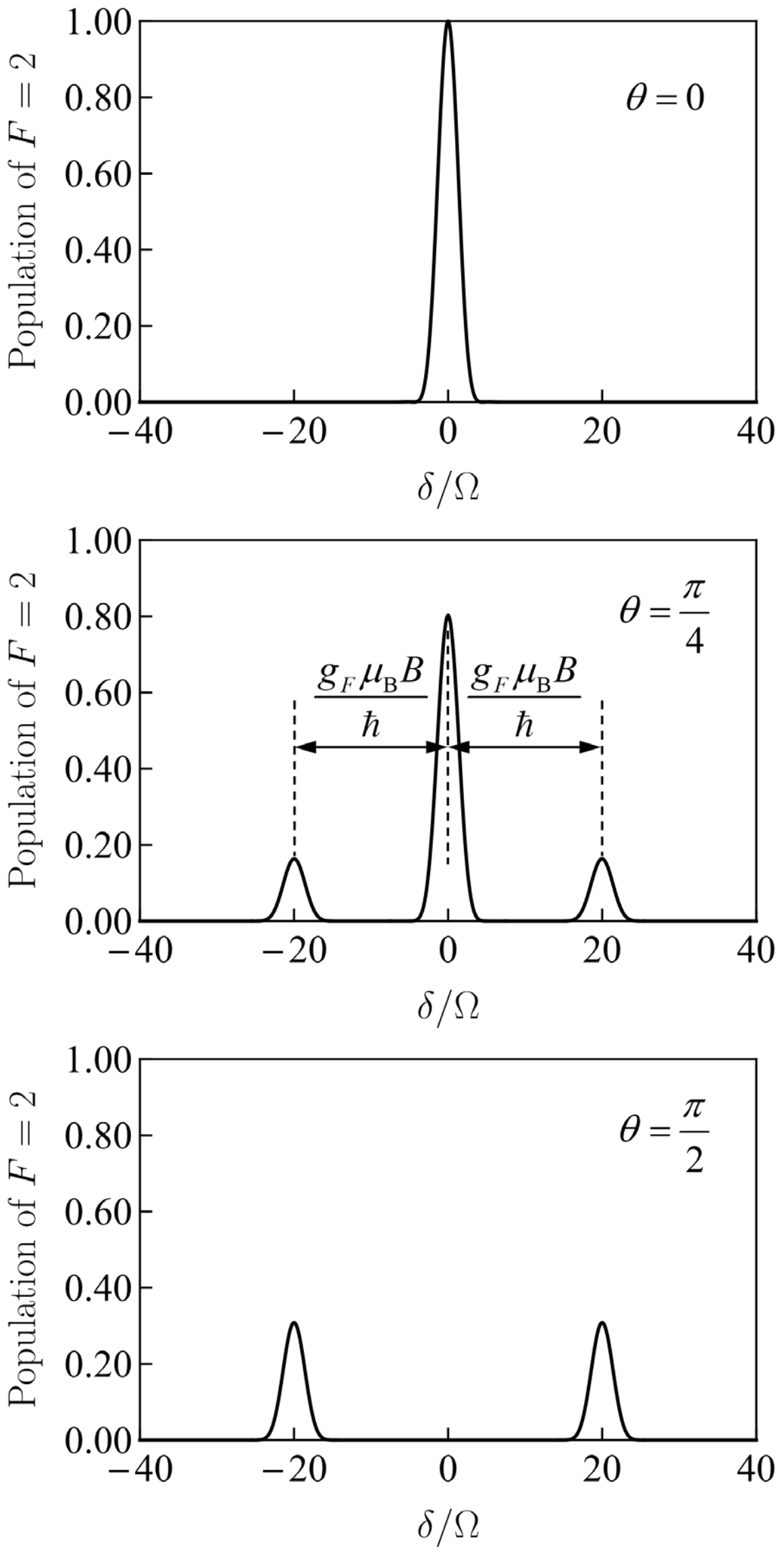

Figure 2. Population of state $F = 2$ after the application of the Blackman microwave pulse when the angle between the magnetic field and the polarization of the microwave field $\theta$ is 0, $\pi/4$, and $\pi/2$. When $\theta = 0$, the microwave field is $\pi$-polarized, and the atoms will only be transited to $F = 2, m_F = 0$, resulting in one resonant peak. When $\theta = \pi/4$, the microwave field is in a superposition of all three polarizations, resulting in three resonant peaks, separated by a distance of $g_F \mu_B B/\hbar$. When $\theta = \pi/2$, the microwave field is in a superposition of the $\sigma^+$ and the $\sigma^-$ polarization, resulting in two resonant peaks.

The Ramsey scheme can also be applied to this vector atomic magnetometer since coherence is well preserved when atoms are in a pure state. To implement the Ramsey scheme, two microwave pulses with an effective duration of $\pi/2\Omega$ instead of a single microwave pulse are applied. An example of the Ramsey spectrum of the magnetometer is illustrated in Figure 3 (a) and Figure 3(b). In each case, the peak value of the Ramsey spectrum would be the same as that for the single-pulse technique described above, and the periodicity would be given by $T^{-1}$, where $T$ is the time separation between the two pulses. The height of a side peak is given by $h = \{1-\cos[(\pi \sin\theta)/2]\}/2$, if the number of the atoms is normalized to unity. If the number of the atoms is not normalized, the height of the peak is $Nh$ for $N$ atoms under interrogation. In this case, the Ramsey signal for the central fringe can be expressed as $S = Nh(1+\cos\phi)/2$, where $\phi$ is the magnetic-field-induced phase shift: $\phi = g_F \mu_B BT/\hbar$. The corresponding quantum projection noise is given by $\Delta S = \sqrt{N(1+h\cos\phi)(2-h-h\cos\phi)}/2$. In the ideal case, the uncertainty of the phase shift is given by $\Delta\phi = \Delta S/|\partial S/\partial\phi|$. The minimum value of $\Delta\phi$ is $\sqrt{\left[1+(1-h)\left(\sqrt{h+1}-h/2\right)\right]/N}$. Noting that $\Delta\phi$ takes the value of $\sqrt{(2-h)/N}$ at $\phi = \pi/2$, it can be seen that this value differs from the minimum value by only $O\left(h^2/\sqrt{N}\right)$. In addition, if other sources of detection noise are dominant, the minimum measurement uncertainty will be achieved at $\phi = \pi/2$, because the phase gradient $|\partial S/\partial\phi|$ is maximized at $\phi = \pi/2$. Therefore, the magnetometer is assumed to be operated at $\phi = \pi/2$. The corresponding minimum measurable value of $B$ in the ideal case is calculated to be $B_{\min} \simeq 28.4\sqrt{(2-h)/N}\sqrt{(\tau_0/\tau)}$ pT for a Ramsey pulse separation of $T = 0.8$ s (pT: pico-Tesla). Here, $\tau_0$ is the duration of one shot of the measurement, which is assumed to be one second, and $\tau$ is the total measurement time. As a specific example, consider the case where $\theta = \pi/4$, such that $h \approx 0.278$. For $N = 10^8$, and a measurement time of one second, it yields $B_{\min} \simeq 3.73$ fT (fT: femto-Tesla).

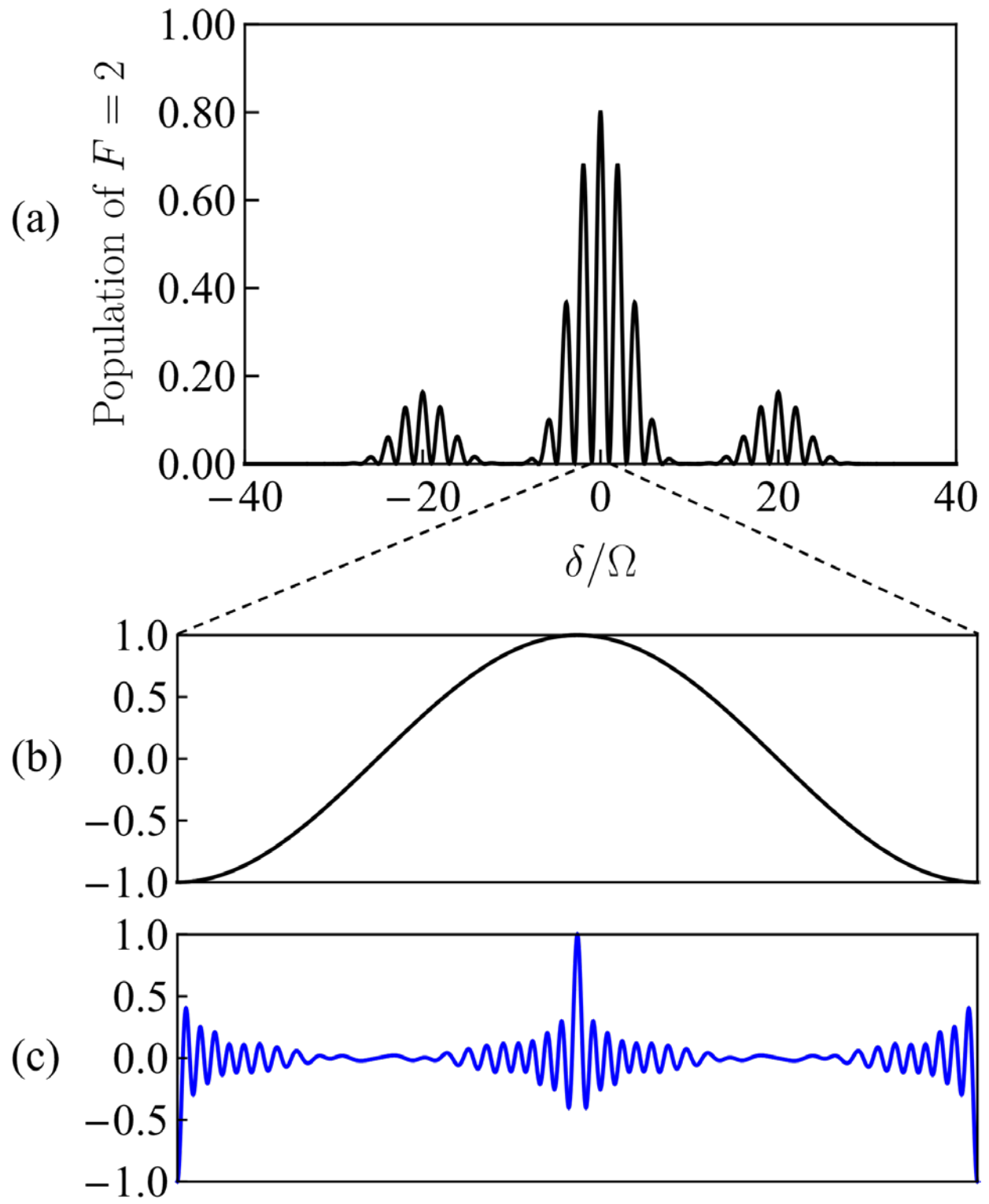

Figure 3. (a) Signal of the magnetometer in a Ramsey configuration where two Blackman microwave pulses are applied for $\theta = \pi/4$. (b) Zoomed-in view of the central fringe in (a). (c) Signal of the magnetometer with the spin squeezing protocol GESP-o for 100 atoms and $\mu = 0.6$ in the plotting range of (b).

The number of atoms initially transferred to the $5S_{1/2}, F = 1, m_F = 0$ state would be a factor in determining the accuracy of the measurement of the magnitude as well as the direction of the magnetic field. In this context, we note that the Rabi frequency of the microwave transition between the $5S_{1/2}, F = 2, m_F = 0$ state and the $5S_{1/2}, F = 1, m_F = 0$ state would depend on $\theta$. As such, the number of transferred atoms can be very small if $\theta$ is close to $\pi/2$. To circumvent this problem, a two-step approach needs to be used. Specifically, we would carry out the whole sequence of measuring $B$ and $\theta$ to establish an initial estimate of these parameters. Given this initial estimate of $\theta$, the experimental sequence will then be repeated by choosing a more proper direction of the polarization and adjusting the duration of the microwave pulse in the state preparation process to maximize the number of atoms transitioned from the $5S_{1/2}, F = 2, m_F = 0$ state to the $5S_{1/2}, F = 1, m_F = 0$ state.

## 3. Spin-squeezed vector atomic magnetometer

Similar to other atomic sensors, like atomic clocks and atom interferometers, spin squeezing could potentially be incorporated to enhance the sensitivity close to the Heisenberg limit. There are two common types of spin squeezing, namely one-axis-twist squeezing (OATS) and two-axis-twist squeezing. Since OATS is simpler and more accessible experimentally, in this paper, we only focus on OATS. In the context of OATS, an atom is modeled as a two-level system, which is equivalent to a spin-1/2 pseudo-spinor mathematically. The Hamiltonian for OATS is $\hbar\chi S_z^2$, where $S_z$ is the $z$-component of the dimensionless collective spin operator. The squeezing parameter, defined as $\mu \equiv \chi t$ ($t$ is the non-linear interaction time), describes the extent of squeezing. As discussed in the previous section, with the method of a preliminary measurement, the magnetometer is completely reduced to a Ramsey atomic clock for each of the three transitions. Given that the application of OATS to a Ramsey clock is discussed in detail in Ref. [22,23,24], it is not difficult to design a similar scheme for the Ramsey magnetometer. The conventional goal of using OATS is to squeeze the quantum noise. However, the suppression of the quantum noise is significant only if the quantum noise level is much higher than the level of other sources of detection noise. Recently, it has been realized that OATS can also be used to magnify the quantum phase shift [22,24,25,26], which will enhance the sensitivity regardless of whether the quantum noise is dominant. The methods for magnifying the quantum phase shift are known as the echo squeezing protocols (ESPs). In these protocols, a squeezing operation as well as the inverse of the squeezing operation are implemented. There are three versions of ESPs, namely the Conventional ESP (CESP), and two types of Generalized ESP (GESP), which are named GESP-e and GESP-o. Briefly, the CESP only magnifies the quantum phase shift, but leaves the quantum noise unchanged compared to atomic sensors without spin squeezing. On the other hand, the GESP not only magnifies the quantum phase shift by a factor of $N \sin \mu / \sqrt{2}$, but also amplifies the quantum noise by a factor of $\sqrt{N} \sin \mu$. Unlike the CESP, the sensitivity of the GESP depends on the parity of $N$ when $\mu$ gets very close to the critical value of $\pi/2$, and GESP-e(o) is optimized for even (odd) values of $N$. The sensitivity of the CESP depends strongly on the value of $\mu$, while the GESP gives a sensitivity of the Heisenberg limit divided by $\sqrt{2}$ for a wide range of values of $\mu$, as shown in Figure 4(a). As such, for the magnetometer, the GESP is preferable since the value of $\mu$ cannot be precisely controlled due to the unknown direction of the magnetic field, which is to be discussed in detail next. We also assume that the GESP would be operated for values of $\mu < \pi / 2$, so that either version of the GESP can be used regardless of the parity of $N$, which is important considering that the parity of a larger number of atoms under interrogation is basically impossible to be determined a priori. An example of the signal of the GESP for $\mu = 0.6$ is shown in Figure 3(c). It can be seen that the fringes are significantly narrowed compared to the signal without spin squeezing. With the GESP, the minimum measurable value of $B$ is reduced by a factor of $\sqrt{N/2}$, so that $B_{\text{min,GESP}} \simeq 40.2\left(\sqrt{2-h}/N\right)\sqrt{(\tau_0/\tau)}$ pT, based on the estimation shown in Sec. 2. If we again consider the case where $\theta = \pi/4$, $N = 10^8$, and a measurement time of one second, it yields $B_{\text{min,GESP}} \simeq 0.527$ aT (aT: atto-Tesla). The minimum measurable magnetic field as a function of $\mu$ is plotted in Figure 4. This sensitivity is higher than that of the best current magnetometers (such as the SQUID magnetometer [27] and the Spin Exchange Relaxation Free atomic magnetometer [8]) by nearly three orders of magnitude.

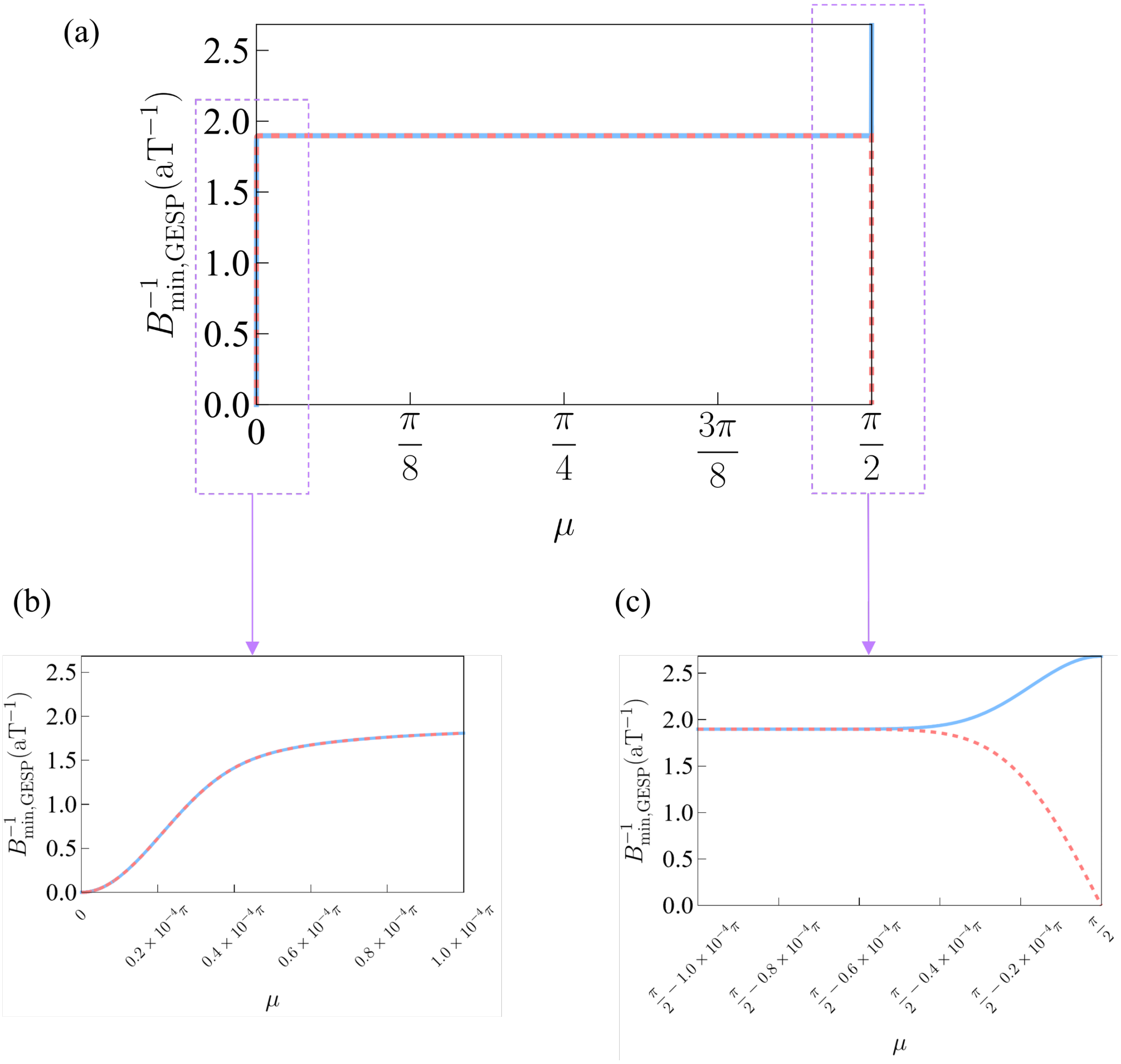

Figure 4. (a) Reciprocal of the minimum measurable magnetic field as a function of $\mu$ that can be achieved by the generalized echo squeezing protocol optimized for even values of $N$ (GESP-e) for $N = 10^8$ (blue solid) and $N = 10^8 + 1$ (red dashed). (b) Zoomed-in view near $\mu = 0$ . (c) Zoomed-in view near $\mu = \pi/2$ .

In what follows, we assume that OATS is realized with cavity-mediated non-linear interaction [28,29,30,31,32]. The non-linear interaction strength can be expressed as $\chi = \delta\bar{n}\varepsilon^2 / \left(\delta^2 + \kappa^2/4\right)$, where $\delta$ is the detuning of the probe from the cavity resonant frequency, $\bar{n}$ is the mean number of photons inside the cavity in the absence of the atoms, $\kappa$ is the linewidth of the cavity, and $\varepsilon$ is the light shift difference between the two ground states involved in the OATS induced by a single photon of the cavity field. Some variations of expression been been derived previously using a detailed analysis of cavity input-output formalism with mean-field approximation and adiabatic elimination of cavity energy levels [28,29]. However, we have found that this expression can be derived in a simpler manner based on solving the eigenvalues and eigenstates of the Hamiltonian, which is presented in the Appendix. Although there is an intuitive interpretation of the mechanism underlying the cavity-mediated non-linear interaction based on a semi-classical model [25], the correct expression for the interaction strength cannot be obtained based on this interpretation. In the Appendix, we also show how to modify this interpretation so that the correct expression can also be obtained.

As illustrated earlier in this section, the magnetometer is identical to a Ramsey atomic clock, for each of the three transitions. As such, one can simply adapt the GESP for an atomic clock to such a magnetometer. It should be noted that for these three transitions, the upper Zeeman substates are different, and thus for the same cavity field, the value of $\chi$ differs for these three transitions. In addition, the optimal cavity resonant frequency that maximizes $\chi$ also differs for these three transitions. As discussed in Sec. 2, to optimize the performance of the magnetometer, the measurement is implemented more than once, with the later measurements making use of the preliminary result obtained from earlier measurements. This technique also applies to the spin-squeezing case. Before implementing spin squeezing, $B$ and $\theta$ can be roughly determined by the unsqueezed magnetometer described in Sec. 2. In this way, the value of $\chi$ for each transition can be calculated for any polarization of the cavity field. However, it should be noted that the cavity field does not only shift the energy level of each Zeeman substate but also couples different Zeeman substates. To minimize the strength of this coupling, a linearly polarized cavity field is preferable [33,34]. To further eliminate the effect of this coupling, it is beneficial to avoid sudden change in the cavity field intensity to ensure adiabatic evolution. Although the preliminary measured $B$ and $\theta$ may not enable a highly precise estimate of $\chi$, a near-Heisenberg-limit sensitivity can still be achieved, since the sensitivity of the GESP does not strongly depend on the value of $\mu$.

## 4. Conclusion

In this paper, we propose a spin-squeezed vector atomic magnetometer employing cold atoms. In this magnetometer, the atoms only undergo coherent processes and thus remain in a pure quantum state, which imposes the requirement of using cold atoms. This magnetometer can enhance the sensitivity in three ways compared to the ones involving relaxation processes. First, the absence of spontaneous emission indicates that only transitions with very narrow natural linewidths are involved in this magnetometer. The narrow linewidth obviously increases the precision in determining the transition frequency. Second, due to the coherent nature of the transitions involved, the Ramsey scheme can be adopted to further narrow the signal fringes. Third, for the same reason, spin squeezing can potentially be integrated to push the sensitivity close to the Heisenberg limit. For experimentally accessible parameters, we show that such a magnetometer could achieve a sensitivity nearly three orders of magnitude higher than that of the best existing magnetometers.

## Acknowledgement:

This work has been supported equally in parts by the Department of Defense Center of Excellence in Advanced Quantum Sensing under Army Research Office grant number W911NF202076, ONR grant number N00014-19-1-2181, and the U.S. Department of Energy, Office of Science, National Quantum Information Science Research Centers, Superconducting Quantum Materials and Systems Center (SQMS) under contract number DE-AC02-07CH11359.

**Appendix**

Here, we derive the expression of the cavity-mediated non-linear coupling coefficient, $\chi$, for one-axis-twist squeezing. The non-Hermitian Hamiltonian [35] for an empty optical cavity can be expressed as (with $\hbar = 1$)

$$\hat{H} = \left(\omega_c - i\frac{\kappa}{2}\right)\hat{a}^\dagger\hat{a} + i\sqrt{\kappa_0}\left(\hat{a}^\dagger\beta e^{-i\omega t} - \hat{a}\beta^* e^{i\omega t}\right) \tag{A1}$$

where $\omega_c$ is the resonant frequency of the cavity, $\omega$ is the frequency of the probe field, as shown in Figure 4(a), $\hat{a}$ is the annihilation operator of the cavity field, $\beta$ is the expectation value of the probe-field annihilation operator, as shown in Figure 4(a). It should be noted that because the probe field is unbounded, the annihilation operator is defined in such a way that $|\beta|^2$ is the number of photons hitting the input coupler mirror in a unit time, which equals the power of the probe field divided by $\hbar\omega$. In this model, the cavity field is described by the quantum model while the probe field is treated classically. In Eq. (A1), $\kappa$ is the linewidth of the cavity, which is determined by the total decay rate, and $\kappa_0$ is the cavity decay rate caused by light transmission through the input coupler mirror. $\kappa$ equals $\kappa_0$ plus the contributions of other sources of loss.

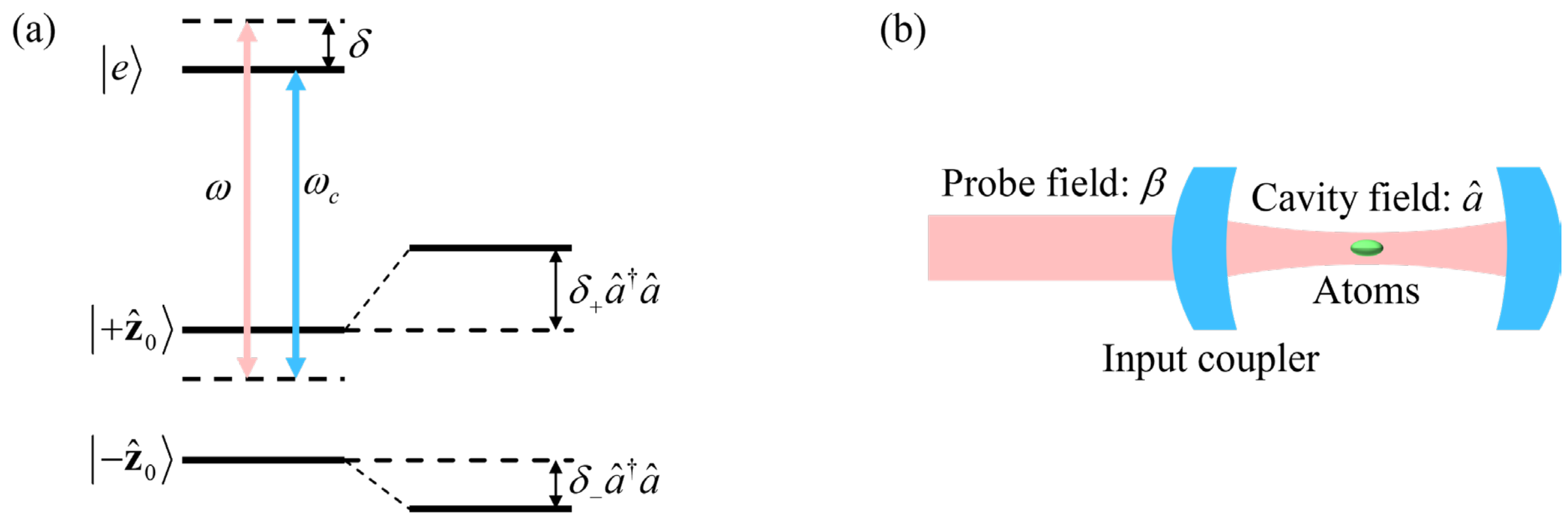

Figure 5. (a) Energy level diagram of an atom model. An atom is modeled as a three-level system consisting of an excited state ($|e\rangle$) and the two ground states ($|+\hat{\mathbf{z}}_0\rangle$, $|-\hat{\mathbf{z}}_0\rangle$). The cavity resonant frequency $\omega_c$ is tuned to a value between the frequencies of the transitions from $|+\hat{\mathbf{z}}_0\rangle, |-\hat{\mathbf{z}}_0\rangle$ to $|e\rangle$. The probe field frequency $\omega$ is detuned from the cavity resonant frequency by $\delta$. The light shift of the state $|\pm\hat{\mathbf{z}}_0\rangle$ induced by a single photon of the cavity field is denoted as $\delta_\pm$. (b) Schematic diagram of the optical cavity for one-axis-twist squeezing. The annihilation operator of the cavity field is denoted as $\hat{a}$, and the expectation value of the probe-field annihilation operator is denoted as $\beta$.

To interpret the meaning of the Hamiltonian, it is helpful to look at the Heisenberg equation given by this Hamiltonian:

$$\dot{\hat{a}} = -i\omega_c \hat{a} - \frac{\kappa}{2}\hat{a} + \sqrt{\kappa_0}\beta e^{-i\omega t} \tag{A2}$$

The first term on the right side of Eq. (A2) describes the oscillation of the cavity field, the second term describes the decay of the cavity field, and the third term describes the probe field transmitted into the cavity. Through the rotating frame transformation $\hat{a} \to \hat{a}e^{-i\omega t}$ , the following time-independent Hamiltonian is obtained:

$$\hat{H} = \left(-\delta - i\frac{\kappa}{2}\right)\hat{a}^\dagger \hat{a} + i\sqrt{\kappa_0}\left(\hat{a}^\dagger \beta - \hat{a}\beta^*\right) \tag{A3}$$

where $\delta \equiv \omega - \omega_c$ . Noting that adding a constant to the Hamiltonian does not change the Heisenberg equation describing the evolution of the system, for the convenience of subsequent steps, the constant $i\sqrt{\kappa_0}\alpha\beta^*$ , where $\alpha = i\sqrt{\kappa_0}\beta/(\delta + i\kappa/2)$, is added to the Hamiltonian, which yields

$$\hat{H} = \left(-\delta - i\frac{\kappa}{2}\right)\hat{a}^\dagger \hat{a} + i\sqrt{\kappa_0}\left(\hat{a}^\dagger \beta - \hat{a}\beta^*\right) + i\sqrt{\kappa_0}\alpha\beta^* \tag{A4}$$

An eigenstate of this Hamiltonian is the coherent state $|\alpha\rangle$ that satisfies the relation $\hat{a}|\alpha\rangle = \alpha|\alpha\rangle$, with the corresponding eigenvalue being zero. This can be proven as follows:

$$\hat{H}|\alpha\rangle = \left[\left(-\delta - i\frac{\kappa}{2}\right)\alpha + i\sqrt{\kappa_0}\beta\right]\hat{a}^\dagger|\alpha\rangle - i\sqrt{\kappa_0}\beta^*\left(\hat{a} - \alpha\right)|\alpha\rangle = 0 \tag{A5}$$

A general expression for the eigenstates of the Hamiltonian (unnormalized) is $|\alpha, n'\rangle = \left(\hat{a}^\dagger - \tilde{\alpha}^*\right)^{n'}|\alpha\rangle$, with the corresponding eigenvalue being $n'(-\delta - i\kappa/2)$, where $n' \in \mathbb{Z}_{\geq 0}$ and $\tilde{\alpha} = i\sqrt{\kappa_0}\beta/(\delta - i\kappa/2)$ . It can be easily seen that $|\alpha, 0\rangle = |\alpha\rangle$ . These eigenstates and eigenvalues can be proven as follows. We first split the Hamiltonian into two parts: $\hat{H}_1 = \left(-\delta - i\frac{\kappa}{2}\right)\hat{a}^\dagger \hat{a} + i\sqrt{\kappa_0}\hat{a}^\dagger\beta$ and $\hat{H}_2 = i\sqrt{\kappa_0}\beta^*\left(\alpha - \hat{a}\right)$, where $\hat{H}_1$ consists of terms including $\hat{a}^\dagger$ and $\hat{H}_2$ consists of the remaining terms. Acting these two parts of Hamiltonian on $|\alpha, n'\rangle$, it

yields

$$\begin{aligned}\hat{H}_1|\alpha,n'\rangle &= \left(-\delta-i\frac{\kappa}{2}\right)\hat{a}^\dagger\left(\hat{a}^\dagger-\tilde{\alpha}^*\right)^{n'-1}\left[\left(\hat{a}^\dagger-\tilde{\alpha}^*\right)\hat{a}+n'\right]|\alpha\rangle+i\sqrt{\kappa_0}\beta\hat{a}^\dagger\left(\hat{a}^\dagger-\tilde{\alpha}^*\right)^{n'}|\alpha\rangle \\ &= \left[\left(-\delta-i\frac{\kappa}{2}\right)\alpha+i\sqrt{\kappa_0}\beta\right]\hat{a}^\dagger\left(\hat{a}^\dagger-\tilde{\alpha}^*\right)^{n'}|\alpha\rangle+n'\left(-\delta-i\frac{\kappa}{2}\right)\hat{a}^\dagger\left(\hat{a}^\dagger-\tilde{\alpha}^*\right)^{n'-1}|\alpha\rangle \\ &= n'\left(-\delta-i\frac{\kappa}{2}\right)\hat{a}^\dagger\left(\hat{a}^\dagger-\tilde{\alpha}^*\right)^{n'-1}|\alpha\rangle\end{aligned} \quad \text{(A6)}$$

$$\begin{aligned}\hat{H}_2|\alpha,n'\rangle &= i\sqrt{\kappa_0}\beta^*\left(\hat{a}^\dagger-\tilde{\alpha}^*\right)^{n'-1}\left[\left(\hat{a}^\dagger-\tilde{\alpha}^*\right)\left(\alpha-\hat{a}\right)-n'\right]|\alpha\rangle \\ &= -i\sqrt{\kappa_0}\beta^* n'\left(\hat{a}^\dagger-\tilde{\alpha}^*\right)^{n'-1}|\alpha\rangle\end{aligned} \quad \text{(A7)}$$

Adding Eq. (A6) and (A7), it yields

$$\begin{aligned}\hat{H}|\alpha,n'\rangle &= n'\left[\left(-\delta-i\frac{\kappa}{2}\right)\hat{a}^\dagger-i\sqrt{\kappa_0}\beta^*\right]\left(\hat{a}^\dagger-\tilde{\alpha}^*\right)^{n'-1}|\alpha\rangle \\ &= n'\left(-\delta-i\frac{\kappa}{2}\right)\left(\hat{a}^\dagger-\tilde{\alpha}^*\right)^{q}|\alpha\rangle = n'\left(-\delta-i\frac{\kappa}{2}\right)|\alpha,n'\rangle\end{aligned} \quad \text{(A8)}$$

which is the eigen equation for the Hamiltonian. From the eigenvalues it can be seen that $|\alpha,n'\rangle$ decays at a rate of $n'\kappa$, and thus only the lowest eigenstate $|\alpha,0\rangle=|\alpha\rangle$ is stable. Therefore, the system will eventually fall into the coherent state $|\alpha\rangle$ regardless of the initial quantum state.

We next consider the case where atoms are exposed to the cavity field. An atom is modeled as a three-level system comprising two ground states and an excited state. When the atom is exposed to a largely detuned light field, the excited state can be adiabatically eliminated, and the net effect of the light field is only the light shifts of the two ground states. Such a two-level system is mathematically equivalent to a spin-1/2 pseudo-spinor. The light shifts of the two ground states are denoted as $\delta_+\hat{a}^\dagger\hat{a}$ and $\delta_-\hat{a}^\dagger\hat{a}$, considering that they are proportional to the intensity of the cavity field. Here, $\delta_\pm$ is the light shift induced by a single photon of the cavity field [positive (negative) for a blue (red) shift], as shown in Figure 4(a). Therefore, the Hamiltonian for the light shifts induced by the cavity light field can be expressed as $\left(\delta_0+\varepsilon\hat{s}_z\right)\hat{a}^\dagger\hat{a}$, where $\delta_0\equiv\left(\delta_++\delta_-\right)/2$, $\varepsilon\equiv\delta_+-\delta_-$, and $\hat{s}_z$ is the *z* component of the dimensionless spin operator of the pseudo-spinor. If an ensemble of atoms is placed in the cavity field, the Hamiltonian for the system can be expressed as

$$\begin{aligned}\hat{H} &= \left(-\delta + \delta_0 + \varepsilon \hat{S}_z - i\frac{\kappa}{2}\right)\hat{a}^\dagger \hat{a} + i\sqrt{\kappa_0}\left(\hat{a}^\dagger \beta - \hat{a}\beta^*\right) + i\sqrt{\kappa_0}\alpha\beta^* \\ &\equiv \left(-\delta' + \varepsilon \hat{S}_z - i\frac{\kappa}{2}\right)\hat{a}^\dagger \hat{a} + i\sqrt{\kappa_0}\left(\hat{a}^\dagger \beta - \hat{a}\beta^*\right) + i\sqrt{\kappa_0}\alpha\beta^*\end{aligned} \tag{A9}$$

It should be noted that in Eq. (A9) the collective spin operator $\hat{S}_z$ instead of the single-atom operator $\hat{s}_z$ is used. In this case, the stable eigenstate of the Hamiltonian is the coherent state $|\alpha'\rangle$, with the eigenvalue being $i\sqrt{\kappa_0}\beta^*(\alpha - \alpha')$, where $\alpha' = i\sqrt{\kappa_0}\beta \big/ \left(\delta' - \varepsilon\hat{S}_z + i\kappa/2\right)$. It should be noted that in the presence of the atoms, this eigenvalue becomes an operator acting on the Hilbert subspace of the atoms. As discussed above, the system will eventually fall into this stable eigenstate. Accordingly, this eigenvalue-based operator can serve as the effective Hamiltonian describing the evolution of the atoms:

$$\hat{H} = i\sqrt{\kappa_0}\beta^*(\alpha - \alpha') \tag{A10}$$

The Hamiltonian shown in Eq. (A10) is already the effective Hamiltonian describing the cavity-mediated non-linear interaction. The following steps are only expanding this Hamiltonian into a Maclaurin series of $\hat{S}_z$.

Before performing the expansion, we first provide some discussion about this Hamiltonian. This effective Hamiltonian can also be expressed as $|\alpha'|^2\left(\delta_0 + \varepsilon\hat{S}_z\right) - \left(|\alpha'|^2 - |\alpha|^2\right)(\delta + i\kappa/2)$. It comprises two parts. The first part $|\alpha'|^2\left(\delta_0 + \varepsilon\hat{S}_z\right)$ is the Hamiltonian describing the light shift of the atoms induced by the cavity field with the intensity corresponding to a mean photon number of $|\alpha'|^2$. The second part $\left(|\alpha'|^2 - |\alpha|^2\right)(-\delta - i\kappa/2)$ describes the change in the cavity field energy resulting from the atoms. From this perspective, the effective Hamiltonian describing the cavity-mediated non-linear interaction can be obtained with a semi-classical model. It should be noted that some qualitative explanation based on a semi-classical model only takes into account the effect described by the first term, namely the state of an atom affects the cavity field intensity, which in turn impacts the evolution of all the atoms [25]. The neglection of the second term will result in a discrepancy of a factor of two if quantitative analysis is implemented based on this explanation.

Now we return to the calculation of the effective Hamiltonian. Expanding the effective Hamiltonian in a Maclaurin series of $\hat{S}_z$, it yields

$$\hat{H} = i\sqrt{\kappa_0}\left(\alpha - \alpha'\right)\beta^* = i\sqrt{\kappa_0}\alpha\beta^*\left(1 - \frac{1}{1 - \varepsilon\hat{S}_z/\left(\delta' + i\kappa/2\right)}\right) = -i\sqrt{\kappa_0}\alpha\beta^*\sum_{k=1}^{\infty}\left(\frac{\varepsilon\hat{S}_z}{\delta' + i\kappa/2}\right)^k \quad \text{(A11)}$$

Keeping only the linear and the quadratic term, the effective Hamiltonian can be written as

$$\begin{aligned}\hat{H} &= |\alpha|^2 \varepsilon\hat{S}_z + \frac{|\alpha|^2 \varepsilon^2\hat{S}_z^2}{\delta' + i\kappa/2} = |\alpha|^2 \varepsilon S_z + \frac{\delta'|\alpha|^2 \varepsilon^2\hat{S}_z^2}{\delta'^2 + \kappa^2/4} - i\frac{\kappa}{2}\frac{|\alpha|^2 \varepsilon^2\hat{S}_z^2}{\delta'^2 + \kappa^2/4} \\ &\equiv |\alpha|^2 \varepsilon\hat{S}_z + \chi\hat{S}_z^2 - \frac{i}{2}\hat{L}^\dagger\hat{L}\end{aligned} \quad \text{(A12)}$$

In can be seen that $\chi = \left[\delta'/\left(\delta'^2 + \kappa^2/4\right)\right]\varepsilon^2|\alpha|^2$ and the Lindblad operator $\hat{L} = \hat{S}_z\sqrt{\chi\kappa/\delta'}$. The Lindblad operator [36] describes the decoherence caused by cavity decay in the squeezing process. Noting that $|\alpha|^2 = \bar{n}$ is the mean number of photons in the cavity field in the absence of atoms and assuming $\delta_0 = 0$ for simplicity, the expression of $\chi$ shown in Sec. 3 is obtained.

---